\documentclass{elsart}
\usepackage{epsfig}
\usepackage{amsmath}
\usepackage{amssymb}

\begin{document}
\begin{frontmatter}

\title{Exact Bright and Dark Spatial Soliton Solutions in Saturable Nonlinear Media}

\author{Gabriel F. Calvo, Juan Belmonte-Beitia, and V\'{\i}ctor M. P\'{e}rez-Garc\'{\i}a}
\address{Departamento de Matem\'aticas, E. T. S. de Ingenieros Industriales and Instituto de Matem\'atica Aplicada a la Ciencia y la Ingenier\'{\i}a (IMACI), \\ E. T. S. I. Industriales, Avda. Camilo Jos\'e Cela, 3 \\ Universidad de Castilla-La Mancha 13071 Ciudad Real, Spain.}
\date{\today}

\begin{abstract}
We present exact analytical bright and dark (black and grey) solitary wave solutions of a nonlinear Schr\"{o}\-din\-ger-type equation describing the propagation of spatial beams in media exhibiting a saturable nonlinearity (such as centrosymmetric photorefractive materials). A qualitative study of the stationary equation is carried out together with a discussion of the stability of the solutions.
\end{abstract}

\end{frontmatter}

\section{Introduction}

The Nonlinear Schr\"{o}dinger Equation (NLSE) in its many versions is one of the most important models in mathematical physics, with applications to numerous fields~\cite{Vazquez,Peyrard} such as, for example, in semiconductor electronics~\cite{Soler,Soler2}, photonics~\cite{Kivshar,Hasegawa}, plasma physics~\cite{Dodd}, fundamentals of quantum mechanics~\cite{fundamentals}, dynamics of accelerators~\cite{Fedele}, mean-field theory of Bose-Einstein condensates~\cite{Dalfovo} or biomolecule dynamics~\cite{Davidov}, to name only a few of them. In some of these fields and many others, the NLSE appears as an asymptotic limit for a slowly varying dispersive wave envelope evolving in a nonlinear medium~\cite{Scott}. The study of these equations has served as a catalyzer for the development of new ideas or even mathematical concepts such as solitons~\cite{Zakharov} or singularities in partial differential equations~\cite{Sulembook,SIAMFibich}. 
\par
Among the many different variants of the nonlinear Schr\"odinger equation, one which has attracted a lot of interest is the so-called local saturable nonlinear Schr\"odinger equation, which we write in the form
\begin{equation}
i\frac{\partial A}{\partial z} + \frac{1}{2}\frac{\partial^{2} A}{\partial x^{2}} + \sigma V(\vert A\vert^{2})A = 0\, , \label{eq:NLPWE}
\end{equation}
where $A$ is a complex envelope function, and $\sigma V(\vert A\vert^{2})$ represents the nonlinearity.  Equation~(\ref{eq:NLPWE}) describes the propagation of a laser beam in a saturable nonlinear medium in  (1+1) dimensions.  $A(x,z)$ measures the (complex) beam amplitude with $x$ and $z$ being the transverse and longitudinal coordinates, respectively. Here, $V(\vert A\vert^{2})$ includes the light-induced refractive index change and $\sigma\in\mathbb{R}$ is proportional to the strength of the nonlinearity. For local saturable nonlinearities, $V(\vert A\vert^{2})\propto (1 +\vert A\vert^{2})^{-\alpha}$, with $\alpha>0$ being the {\em saturation index}. Equation~(\ref{eq:NLPWE}) has shown a number of applications in plasma physics~\cite{Yu} and nonlinear optics~\cite{Marburger,Herrmann,Krolikowski93}, specifically in photorefractive media. Photorefractive materials manifest a wealth of nonlinear phenomena which include the propagation of solitons~\cite{Segev94,Christodoulides95,Segev96,Krolikowski00,Calvo02}, surface waves~\cite{Quirino} and slow light~\cite{Shumelyuk}, pattern formation~\cite{Mamaev}, charge singularities~\cite{Calvo00}, and critical enhancement~\cite{Podivilov}. The case with $\alpha=1$ (which corresponds to non-centrosymmetric photorefractive media) has been throughly studied, and solitary wave solutions in the form of bright, black and grey solitons have been found numerically~\cite{Segev94,Christodoulides95,Segev96,Carvalho07}. It was also shown theoretically (by means of numerical calculations)~\cite{SegevAgranat}, and experimentally~\cite{KLTN1,KLTN2}, that the nonlinearity with $\alpha=2$ also appears in the, so-called, centrosymmetric photorefractive media and exhibits bright and black spatial solitons. Motivated by these studies, here we go beyond the previous numerical studies and obtain three classes of exact solutions to Eq. (\ref{eq:NLPWE}) in closed form: bright, black and grey solitons (the latter class being addressed here for the first time, to the best of our knowledge).
\par
Although many methods have been used to construct exact solutions to various types of nonlinear Schr\"odinger equations such as the cubic one in different scenarios~\cite{CSF1,CSF2,Nuestro,Nuestro2}, the  cubic-quintic~\cite{CSF3,CSF4} and others~\cite{JVV}, it is remarkable that no exact solutions are available for the  saturable nonlinear Schr\"odinger equation~(\ref{eq:NLPWE}) for $\alpha=2$ despite its practical applications.
\par
The paper is organized as follows. In section 2, we first impose the necessary conditions to derive solitary waves of a general version of the nonlinear paraxial wave equation in which the nonlinearity depends locally on the light intensity (the latter being proportional to $\vert A\vert^{2}$). Section 3 presents a  qualitative study of the stationary saturable nonlinear Schr\"odinger equation. Sections 4, and 5, contain the main results about the study of analytical solutions to Eq. (\ref{eq:NLPWE}): we obtain exact bright, and dark solitary waves, respectively, together with their existence curves, that constrain the amplitude and the width of the waves. Stability of bright and dark solutions is also discussed.
\par
\section{Solitary Waves of the Nonlinear Pa\-ra\-xial Wa\-ve Equation}

For completeness, our first purpose is to deal with a general local nonlinearity $\sigma V(\vert A\vert^{2})$ to derive the characteristic wave equation for solitary waves from Eq.~(\ref{eq:NLPWE}). It is more convenient to express the solutions of Eq.~(\ref{eq:NLPWE}) in terms of the real amplitude $u(x,z)$ and phase $\phi(x,z)$  as
\begin{equation}
A(x,z) = u(x,z)\, \textrm{e}^{i\phi(x,z)}\, . \label{eq:sol1}
\end{equation}
Substitution of Eq.~(\ref{eq:sol1}) into (\ref{eq:NLPWE}) yields the system of equations
\begin{subequations}
\begin{eqnarray}
\frac{\partial^{2} \phi}{\partial x^{2}} + 2\frac{\partial\ln u}{\partial x}\frac{\partial \phi}{\partial x} + 2\frac{\partial\ln u}{\partial z} &=& 0\, , \label{eq:ac1}\\
\frac{\partial \phi}{\partial z} + \frac{1}{2}\left[ \left( \frac{\partial \phi}{\partial x}\right)^{2} - \frac{1}{u}\frac{\partial^{2} u}{\partial x^{2}}\right] - \sigma V(u^{2}) &=& 0\, . \label{eq:ac2}
\end{eqnarray}  
\end{subequations}
It is clear that solving the coupled system of Eqs.~(\ref{eq:ac1}) and (\ref{eq:ac2}) is equivalent to solving Eq.~(\ref{eq:NLPWE}).
\par
We now impose the fundamental property of solitary waves: their shape invariance with respect to Galilean boosts
\begin{equation}
u(x,z) = u(x - vz)\, , \label{eq:galilean}
\end{equation}
where $v$ can be regarded as the dimensionless {\em transverse spatial velocity} (or the steering angle between the propagation direction and the $z-$axis). In particular, when $v=0$, the profile of solitary waves remains unchanged with respect to translations along the $z-$axis. Let $\xi = x - vz$ and $\zeta = z$. Then, the system of Eqs.~(\ref{eq:ac1}) and (\ref{eq:ac2}) reads as
\begin{subequations}
\begin{eqnarray}
\frac{\partial^{2} \phi}{\partial \xi^{2}} + 2\frac{\partial\ln u}{\partial \xi}\frac{\partial \phi}{\partial \xi} - 2v\frac{\partial\ln u}{\partial \xi}  &=& 0\, , \label{eq:bc1}\\
\frac{\partial \phi}{\partial \zeta} -v\frac{\partial \phi}{\partial \xi} + \frac{1}{2}\left[ \left( \frac{\partial \phi}{\partial \xi}\right)^{2} - \frac{1}{u}\frac{\partial^{2} u}{\partial \xi^{2}}\right] - \sigma V(u^{2}) &=& 0\, . \label{eq:bc2}
\end{eqnarray} 
\end{subequations}
Equation (\ref{eq:bc1}) can be readily integrated with respect to $\xi$, and yields
\begin{equation}
\phi(\xi,\zeta) = C(\zeta) + v\xi + D(\zeta)\int^{\xi}\frac{d\tau}{u^{2}(\tau)}\, , \label{eq:phi2}
\end{equation}
where $C(\zeta)$ and $D(\zeta)$ are two integration functions that depend solely on $\zeta$. Also, since $\phi(\xi,\zeta)$ satisfies Eq.~(\ref{eq:bc2}), inserting expression (\ref{eq:phi2}) into Eq.~(\ref{eq:bc2}) one finds
\begin{equation}
2k\frac{\partial}{\partial \zeta} \left[ C(\zeta) + D(\zeta)\int^{\xi}\frac{d\tau}{u^{2}(\tau)} \right] - v^{2} + \frac{D^{2}(\zeta)}{u^{4}} = \frac{1}{u}\frac{\partial^{2} u}{\partial\xi^{2}} + 2\sigma V(u^{2}) , \label{eq:bc3}
\end{equation}
that must be fulfilled for all $\xi$ and $\zeta$. Now, if $u(\xi)$ is a non-constant function, this implies that the only admissible solutions for $C(\zeta)$ and $D(\zeta)$ are
\begin{subequations}
\begin{eqnarray}
C(\zeta) & = & \Gamma_{0} \zeta + \phi_{0} \, , \label{eq:cz}\\
D(\zeta) & = & l_{0} \, , \label{eq:dz}
\end{eqnarray}
\end{subequations}
where $\Gamma_{0}$ is a constant (propagation constant), $\phi_{0}$ is a fixed reference phase, and $l_{0}$ is a parameter. Under these conditions, Eq.~(\ref{eq:bc3}) reduces to the following {\em characteristic equation for solitary waves}
\begin{equation}
\frac{d^{2} u}{d\xi^{2}} - 2\Gamma u - \frac{l_{0}^{2}}{u^{3}} + 2\sigma V(u^{2})u  = 0\, , \label{eq:bc4}
\end{equation}
with $\Gamma\equiv\Gamma_{0} - \frac{v^{2}}{2}$. If $l_{0}\neq0$, in order to obtain nonsingular solutions to Eq.~(\ref{eq:bc4}), it is necessary that $u\neq 0$ for $-\infty<\xi<\infty$. Multiplying  Eq. (\ref{eq:bc4}) by $du/d\xi$ and upon integration, it follows that
\begin{equation}
\frac{1}{2}\left( \frac{du}{d\xi}\right)^{2}  - \Gamma u^{2} + \frac{l_{0}^{2}}{2u^{2}} + \sigma U(u)= \mathcal{E}_{0}\, , \label{eq:csweqint}
\end{equation}
where $U(u) \equiv 2\int^{u}V(\tau^{2})\tau d\tau$ and $\mathcal{E}_{0}$ is a constant (it plays the role of an effective total mechanical energy). It is expected that if the nonlinear contribution is absent, i.e. if $\sigma U(u)=0$, then, the solution to Eq. (\ref{eq:csweqint}), given by 
\begin{equation}
u^{2}(\xi)=\frac{1}{\sqrt{2\Gamma}}\left\{ \frac{1}{4} \left[ \textrm{e}^{\pm \sqrt{2\Gamma}(\xi - \xi_{0})} -\sqrt{\frac{2}{\Gamma}}\,\mathcal{E}_{0}\, \textrm{e}^{\mp \sqrt{2\Gamma}(\xi - \xi_{0})}\right]^{2} + l_{0}^{2}\,\textrm{e}^{\mp 2\sqrt{2\Gamma}(\xi - \xi_{0})}\right\}\! ,\nonumber\\
\label{eq:Unl0}
\end{equation}
cannot represent a localized wave for any value of $\Gamma$, $l_{0}$, and $\mathcal{E}_{0}$. The presence of $\sigma U(u)$ is thus essential for the existence of solitary waves. Henceforth, we assume $\sigma\neq0$. Moreover, for solitary waves to exist it is clearly required that $\mathcal{E}_{0}-\sigma U(u) + \Gamma u^{2} - l_{0}^{2}/(2u^{2})\geq0$.
\par
Since spatial solitary waves are well-localized excitations in space, we look for solutions such that their derivatives of all order vanish at infinity. Three families of solutions are distinguished according to additional boundary conditions at infinity and at the origin $\xi=\xi_{0}$. Let $u =  u_{\infty}$ for $|\xi|\rightarrow \infty$, and $u =  u_{0}$, $\frac{du}{d\xi} = u_{0}'$ both at $\xi = \xi_{0}$. Solitary waves whose asymptotics tend to zero ($u_{\infty}=0$), together with $u_{0}'=0$, are the well-known {\em bright} solitary waves. Those that fulfill $u_{0}= 0$ (with $u_{\infty}\neq0$ and $u_{0}'\neq0$) are the {\em black} solitary waves, whereas those that do not vanish in $-\infty<\xi<\infty$ and satisfy $u_{0}'=0$ are {\em grey} solitary waves (both black and grey are denoted as \emph{dark} solitons). In contrast with bright and black solutions, grey solitary waves possess nonzero values of $l_{0}$. Using the above prescribed boundary conditions, formulae for the constants $\Gamma$, $l_{0}$, and $\mathcal{E}_{0}$ can be found for each family of solutions. 
\par
Bright solitary waves correspond to $0\leq u^{2}\leq u_{0}^{2}$,
\begin{equation}
\Gamma =  \frac{\sigma\left[U(u_{0})-U(0)\right]}{u_{0}^{2}}\, , \quad l_{0}=0\, , \quad \mathcal{E}_{0} = \sigma U(0)\, ,\label{eq:brightpropgconst}
\end{equation}
and are the implicit solutions obtained after integration of
\begin{equation}
\frac{1}{2\sigma}\left(\frac{du}{d\xi}\right)^{2}= U(0)-U(u) +\frac{\left[U(u_{0})-U(0)\right] u^{2}}{u_{0}^{2}} \, .
\label{eq:brightsolitarywaves}
\end{equation}
Black solitary waves correspond to $0\leq u^{2}\leq u_{\infty}^{2}$,
\begin{equation}
\Gamma =  \sigma V(u_{\infty}^{2})\, ,  \quad l_{0}=0\, , \quad \mathcal{E}_{0} = \sigma \left[ U(u_{\infty})-U(u) - u_{\infty}^{2} V(u_{\infty}^{2})\right]\, ,\label{eq:darkpropgconst}
\end{equation}
and obey
\begin{equation}
\frac{1}{2\sigma}\left(\frac{du}{d\xi}\right)^{2}= U(u_{\infty})-U(u) - \left(u_{\infty}^{2} - u^{2}\right) V(u_{\infty}^{2})\, .
\label{eq:darksolitarywaves}
\end{equation}
Finally, grey solitary waves correspond to $0<\textrm{min}\{u_{0}^{2},u_{\infty}^{2}\}\leq u^{2}\leq \textrm{max}\{u_{0}^{2},u_{\infty}^{2}\}$,

\begin{subequations}
\begin{eqnarray}
\Gamma & = &  \frac{\sigma\left\{\left[U(u_{0})-U(u_{\infty})\right]u_{0}^{2}+(u_{\infty}^{2}-u_{0}^{2})V(u_{\infty}^{2})u_{\infty}^{2}\right\}}{(u_{\infty}^{2}-u_{0}^{2})^{2}}\, , \label{eq:greypropgconst}\\
l_{0}^{2} & = & \frac{2\sigma\left[U(u_{\infty})-U(u_{0})-(u_{\infty}^{2}-u_{0}^{2})V(u_{\infty}^{2})\right]u_{0}^{2}u_{\infty}^{4}}{(u_{\infty}^{2}-u_{0}^{2})^{2}}\, , \label{eq:greylo}\\
\mathcal{E}_{0} &=& \frac{\sigma\left[(u_{\infty}^{4}+u_{0}^{4})U(u_{\infty})-(u_{\infty}^{4}-u_{0}^{4})V(u_{\infty}^{2})u_{\infty}^{2}-2u_{0}^{2}\,u_{\infty}^{2}U(u_{0})\right]}{(u_{\infty}^{2}-u_{0}^{2})^{2}}\, , \label{eq:greyEo}
\end{eqnarray}
\end{subequations}
and are governed by 
\begin{eqnarray}
\frac{1}{2\sigma}\left(\frac{du}{d\xi}\right)^{2}&=& \frac{(u_{\infty}^{4}+u_{0}^{4})U(u_{\infty})-(u_{\infty}^{4}-u_{0}^{4})V(u_{\infty}^{2})u_{\infty}^{2}-2u_{0}^{2}\,u_{\infty}^{2}U(u_{0})}{(u_{\infty}^{2}-u_{0}^{2})^{2}}\nonumber \\
&-& U(u) - \frac{\left[U(u_{\infty})-U(u_{0})-(u_{\infty}^{2}-u_{0}^{2})V(u_{\infty}^{2})\right]u_{0}^{2}\,u_{\infty}^{4}}{(u_{\infty}^{2}-u_{0}^{2})^{2}\,u^{2}} \nonumber \\
&+&\frac{\left\{\left[U(u_{0})-U(u_{\infty})\right]u_{0}^{2}+(u_{\infty}^{2}-u_{0}^{2})V(u_{\infty}^{2})u_{\infty}^{2}\right\}\!u^{2}}{(u_{\infty}^{2}-u_{0}^{2})^{2}}\, .
\label{eq:greysolitarywaves}
\end{eqnarray}
When $u_{\infty}=0$ or $u_{0}=0$, one recovers from Eq.~(\ref{eq:greysolitarywaves}) the bright and dark Eqs.~(\ref{eq:brightsolitarywaves}) and~(\ref{eq:darksolitarywaves}), respectively. Moreover, depending on the given forms for $V(u^{2})$ and $U(u)$, together with the particular values for $u_{\infty}$ and $u_{0}$, the nonlinearity coefficient $\sigma$ can be positive or negative. 
\par
In the next sections we carry out the analysis for bright, dark and grey solitary wave families when the nonlinearity is of the saturable form
\begin{equation}
V(u^{2})=\left(\frac{1+u_{\infty}^{2}}{1+u^{2}}\right)^{2}\, ,\label{eq:quadraticsaturable}
\end{equation}
and therefore $U(u)=-(1+u_{\infty}^{2})^{2}/(1+u^{2})$. Notice that the cubic-like nonlinearity (e.g. in Kerr media) corresponds for Eq.~(\ref{eq:quadraticsaturable}) to the limit when $u^2\ll1$; that is, for $V(u^{2})=(1+u_{\infty}^{2})^{2}(1-2u^{2})$ and $U(u)=(1+u_{\infty}^{2})^{2}(u^{2}-u^{4})$.

\section{Qualitative Analysis}

\begin{figure}
\begin{center}
\hspace*{-0.1cm}
\hbox{\vbox{\vskip -0.0cm \epsfig{figure=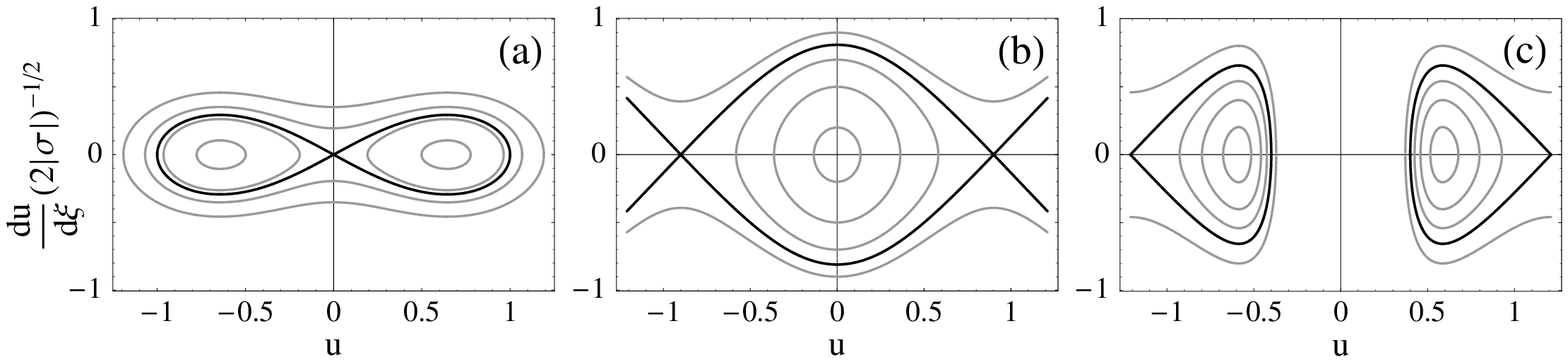, scale=0.61}}}
\end{center}
\vspace*{-0.2cm}
\caption[\bf]{\small Phase-plane portrait of the real solutions to Eq. (\ref{eq:bcgeneral}) for (a) $\sigma=2\Gamma<0$ and $u_{\infty}=0$; (b) $\sigma=\Gamma>0$ and $u_{\infty}=0.9$; (a) and (b) both with $l_{0}=0$. (c) $l_{0}\neq0$, $\sigma=1.2\Gamma>0$ and $u_{\infty}=1.2$.}
\label{fig:bdg}
\end{figure}

Before deriving explicit localized solutions to Eqs.~(\ref{eq:brightsolitarywaves}), (\ref{eq:darksolitarywaves}), and (\ref{eq:greysolitarywaves}) when $V(u^{2})$ is given by~(\ref{eq:quadraticsaturable}), we will carry out a qualitative analysis of Eq.~(\ref{eq:bc4}) with the latter form for $V(u^{2})$:
\begin{equation}
\frac{d^{2} u}{d\xi^{2}} - 2\Gamma u  - \frac{l_{0}^{2}}{u^{3}} + 2\sigma\left(\frac{1+u_{\infty}^{2}}{1+u^{2}}\right)^{2} u  = 0\, . \label{eq:bcgeneral}
\end{equation}
Let us first examine the case $l_{0}=0$. One easily verifies that Eq.~(\ref{eq:bcgeneral}) has three possible equilibrium points
\begin{equation}
u=0\, ,\quad u_{\pm} = \pm\sqrt{(1+u_{\infty}^{2})\sqrt{\frac{\sigma}{\Gamma}}-1} \, ,\label{eq:equilibriumlo0}
\end{equation}
if $(1+u_{\infty}^{2})>\sqrt{\Gamma/\sigma}$. When $\sigma<0$ and $\Gamma<0$, $u=0$ is a saddle point and $u_{\pm}$ are centers [see Fig.~\ref{fig:bdg}(a)], whereas for $\sigma>0$ and $\Gamma>0$, $u=0$ is a center and $u_{\pm}$ are saddle points [see Fig.~\ref{fig:bdg}(b)]. In Fig.~\ref{fig:bdg}(a), the closed orbits of the phase-plane portrait $(u,du/d\xi)$ correspond to periodic solutions of Eq. (\ref{eq:bcgeneral}).  It is apparent that between the external and internal closed orbits there exists a homoclinic orbit (black curve), which, will be identified as a bright soliton solution in the next section. With respect to Fig.~\ref{fig:bdg}(b), we have a heteroclinic orbit (black curve) in the phase-plane portrait, and, as we shall see in section 5, it will correspond to a black soliton solution. Inside this orbit, there exist closed orbits which, again, represent periodic solutions of Eq. (\ref{eq:bcgeneral}). Outside the orbit, the solutions are no longer periodic.
\par
When $l_{0}\neq0$, Eq.~(\ref{eq:bcgeneral}) has a repulsive singularity and possess four nonzero equilibrium points $u_{j}$, $j=1,2,3,4$; they correspond to two centers $u_{1}=-u_{2}$ (we omit their lengthy expressions), and two saddle points $u_{3}=-u_{4}=u_{\infty}$. Two homoclinic orbits appear [see black curves in Fig.~\ref{fig:bdg}(c)], which will be interpreted in section 5 as grey solitary waves. Similarly with the cases in which $l_{0}=0$, the inner closed orbits Fig.~\ref{fig:bdg}(c) are periodic solutions.
\section{Bright Solitary Waves}

\subsection{Explicit expressions for bright solitons}

\begin{figure}
\begin{center}
\hspace*{-0.3cm}
\hbox{\vbox{\vskip -0.0cm \epsfig{figure=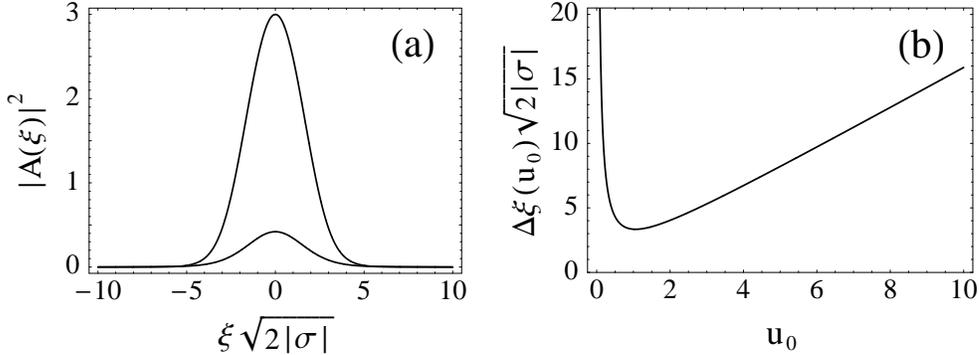, scale=0.870}}}
\end{center}
\vspace*{-0.2cm}
\caption[\bf]{\small (a) Transverse profiles of bright solitary waves for $u_{0}^{2}=3$ and $u_{0}^{2}=0.5$, as obtained from Eq. (\ref{eq:brightcssolution}). (b) Existence curve of bright solitary waves, where the FWHM $\Delta\xi$ is plotted versus the peak amplitude $u_{0}$.}
\label{fig:BrightCentrosymm}
\end{figure}

Fundamental bright solitary waves fulfill $u_{\infty}=u_{0}'=0$, and, without loss of generality, we may assume that $u_{0}> 0$. Using Eqs.~(\ref{eq:brightpropgconst}),~(\ref{eq:brightsolitarywaves}), and~(\ref{eq:quadraticsaturable}), the constant $\Gamma=\sigma/(1+u_{0}^{2})$, and the resulting energy equation is 
\begin{equation}
\left( \frac{du}{d\xi}\right)^{2} = -\frac{2\sigma \,(u_{0}^{2}-u^{2})u^{2}}{(1+u_{0}^{2})(1+u^{2})}\, .\label{eq:csweqint2}
\end{equation}
Now, since $u(\xi)$ is a real function, it is thus necessary that $\sigma<0$. Equation~(\ref{eq:csweqint2}) can be integrated exactly and it yields an implicit relation for the envelope distribution $u(\xi)$ of bright solitary waves 
\begin{equation}
\sqrt{\frac{1+u_{0}^{2}}{2\vert\sigma\vert}}\left\{ \textrm{arctan}\left[ \sqrt{\frac{u_{0}^{2}-u^{2}}{1+u^{2}}}\right] + \frac{1}{u_{0}}\textrm{arctanh}\left[ \sqrt{\frac{u_{0}^{2}-u^{2}}{(1+u^{2})u_{0}^{2}}}\right] \right\} = \pm(\xi - \xi_{0})\,. 
 \label{eq:brightcssolution}
\end{equation}
Notice that, in this case, the obtained solution corresponds to the homoclinic orbit or {\em loop} of Fig. \ref{fig:bdg}(a). Bright solitary profiles given by Eq.~(\ref{eq:brightcssolution}) are shown in Fig.~\ref{fig:BrightCentrosymm}(a). In the limit $u\leq u_{0}\ll 1$ we recover from Eq.~(\ref{eq:brightcssolution}) the well-known profile $u(\xi)=\pm u_{0}\,\textrm{sech}[u_{0}\sqrt{ 2\vert\sigma\vert}\,(\xi - \xi_{0})]$ of the one-dimensional bright soliton in a nonlinear cubic medium. Moreover, the dependence of the full width at half maximum (FWHM), $\Delta\xi$, as a function of the peak amplitude $u_{0}$ can be easily obtained from Eq.~(\ref{eq:brightcssolution}) by setting $u^{2}(\Delta\xi/2)=u_{0}^{2}/2$, and reads
\begin{equation}
\Delta\xi (u_{0})= \sqrt{\frac{2(1+u_{0}^{2})}{\vert\sigma\vert}}\left\{ \textrm{arctan}\left[ \sqrt{\frac{u_{0}^{2}}{2+u_{0}^{2}}}\right] + \frac{1}{u_{0}}\textrm{arctanh}\left[ \frac{1}{\sqrt{2+u_{0}^{2}}}\right] \right\} . 
\label{eq:brightcsexistcurv}
\end{equation}
Figure~\ref{fig:BrightCentrosymm}(b) displays the form of Eq.~(\ref{eq:brightcsexistcurv}). It is also known as the {\em existence curve} for bright solitary waves because it provides a relation between the FWHM, the peak amplitude of the beam and the nonlinearity strength (through the parameter $\sigma$) for which bright solitary waves can exist. Such a dependence is often measured in experiments~\cite{KLTN1}. In our present case there are several interesting features. A property of $\Delta\xi(u_{0})$ is the so-called {\em amplitude bistability}. This means that for a given FWHM there are two different values of $u_{0}$ for which bright solitary waves are to be found. This is depicted in Fig. \ref{fig:BrightCentrosymm}(a), where the two exhibited bright solitons are characterized by the same FWHM. It is evident that with increasing peak intensity, the bright solitary waves have a wider width. This is because the nonlinearity becomes weaker for large $u_{0}$, and, therefore, in order to arrest the broadening (due to diffraction), $\Delta\xi(u_{0})$ has to be larger. All these features are inherent of bright solitons in many saturable nonlinearities and constitute an ingredient for stability of the solutions~\cite{Segev96}. The minimum of the existence curve can be evaluated by setting $\partial\Delta\xi/\partial u_{0} = 0$ in Eq.~(\ref{eq:brightcsexistcurv}), and is obtained for $u_{0}\simeq 1.059$ (which gives $\sqrt{2\vert\sigma\vert}\Delta\xi\simeq 3.338$). These values are in exact agreement with those calculated numerically in Ref.~\cite{SegevAgranat}.
\par

\subsection{Stability}

\begin{figure}
\begin{center}
\hspace*{-0.3cm}
\hbox{\vbox{\vskip -0.0cm \epsfig{figure=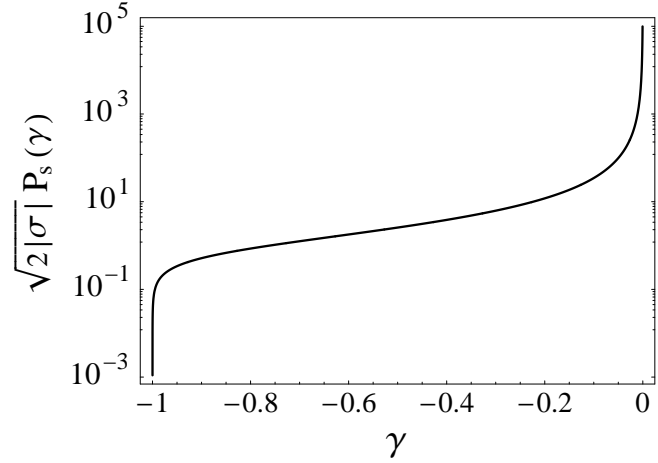, scale=0.9}}}
\end{center}
\vspace*{-0.3cm}
\caption[\bf]{\small Dependence of the power $P_{\textrm{s}}$ on the propagation constant, where $\gamma = \Gamma/\vert\sigma\vert$.}
\label{fig:BrightCentrosymmPower}
\end{figure}

So far, we have been able to develop a full analytical approach allowing us to determine the exact profiles and the existence curve of bright solitary waves. However, an important aspect still remains open: {\em Are the found bright solitary waves linearly stable with respect to small perturbations?} Using a well-known stability criterion for fundamental bright solitary waves~\cite{VK,Laedke}, we answer to this question. All we need to do is to calculate the dependence of the power $P_{\textrm{s}}$ of the bright solitary waves on the propagation constant $\Gamma_{0}$. More explicitly,
\begin{multline}
P_{\textrm{s}}  = \int^{\infty}_{-\infty} u^{2}(x) dx = 2 \int^{\infty}_{0} u^{2}(x) dx = -2 \int^{0}_{u_{0}} \frac{u^{2}du}{\left(\frac{du}{dx}\right)} \\
= \sqrt{\frac{2(1+u_{0}^{2})}{\vert\sigma\vert}}\int^{u_{0}}_{0}\! \sqrt{\frac{1+u^{2}}{u_{0}^{2}-u^{2}}}u^{2} du =\sqrt{\frac{1+u_{0}^{2}}{2\vert\sigma\vert}}\int^{u_{0}^{2}}_{0}\! \sqrt{\frac{1+\eta}{u_{0}^{2}-\eta}}d\eta \, ,
\label{eq:calculationPs}
\end{multline}
where we have employed the symmetry property of the squared amplitude $u^{2}(-\xi)=u^{2}(\xi)$ and Eq.~(\ref{eq:csweqint2}). After integration, the result is
\begin{equation}
P_{\textrm{s}}(u_{0}) = \sqrt{\frac{1+u_{0}^{2}}{2\vert\sigma\vert}}\left[ u_{0} + (1 + u_{0}^{2})\,\textrm{arctan}\, u_{0}\right]  . \label{eq:brightpowerscs}
\end{equation}
Using $\Gamma=\sigma/(1+u_{0}^{2})$, the power $P_{\textrm{s}}$ can be cast in the form
\begin{equation}
P_{\textrm{s}}(\gamma) = \frac{1}{\vert\gamma\vert\sqrt{2\vert\sigma\vert}} \left\{ \sqrt{1-\vert\gamma\vert} + \frac{1}{\sqrt{\vert\gamma\vert}}\, \textrm{arctan}\left[ \sqrt{\frac{1}{\vert\gamma\vert}-1}\right] \right\}  , \label{eq:brightpowerscs2}
\end{equation}
where $\gamma = \Gamma/\vert\sigma\vert$. Fundamental bright solitary waves  are stable if $\partial P_{\textrm{s}}/\partial\Gamma_{0} >0$, and unstable otherwise. Figure~\ref{fig:BrightCentrosymmPower} shows that $P_{\textrm{s}}$ is an increasing function of the propagation constant $\Gamma_{0}$ for $v^{2}/2- \vert\sigma\vert<\Gamma_{0}< v^{2}/2$. Accordingly, the found bright solitary waves are all of them stable.
\par
\section{Dark Solitary Waves}
\subsection{Black Solitary Waves}

Fundamental black solitary waves fulfill $u_{0}= 0$ ($u_{\infty}\neq0$ and $u_{0}'\neq0$). Using Eqs.~(\ref{eq:darkpropgconst}),~(\ref{eq:darksolitarywaves}), and~(\ref{eq:quadraticsaturable}) we find that $\Gamma=\sigma$ and the energy equation is given by
\begin{equation}
\left( \frac{du}{d\xi}\right)^{2} = \frac{2\sigma\,(u_{\infty}^{2}-u^{2})^{2}}{1+u^{2}}\, . \label{eq:darkcsweqint2}
\end{equation}
Again, since $u(\xi)$ is a real function, it is necessary now that $\sigma>0$. 
\par
\begin{figure}
\begin{center}
\hspace*{-0.3cm}
\hbox{\vbox{\vskip -0.0cm \epsfig{figure=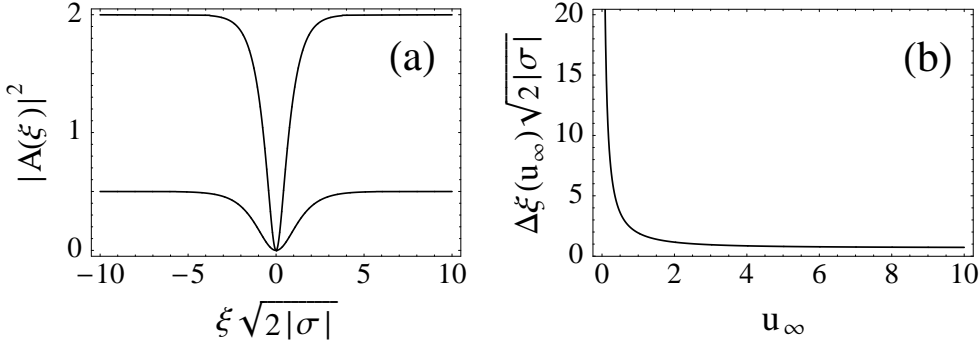, scale=0.91}}}
\end{center}
\vspace*{-0.2cm}
\caption[\bf]{\small (a) Transverse profiles of black solitary waves for $u_{\infty}^{2}=2$ and $u_{\infty}^{2}=1/2$ as obtained from Eq.~(\ref{eq:darkcssolution}). (b) Existence curve of black solitary waves, where the FWHM $\Delta\xi$ is plotted versus the amplitude at infinity $u_{\infty}$.}
\label{fig:DarkCentrosymm}
\end{figure}

Equation~(\ref{eq:darkcsweqint2}) can be integrated exactly and it yields an implicit relation for the envelope distribution of black solitary waves 
\begin{equation}
\frac{\sqrt{1+u_{\infty}^{2}}}{u_{\infty}}\,\textrm{arctanh}\left(\frac{u\,\sqrt{1+u_{\infty}^{2}}}{u_{\infty}\sqrt{1+u^{2}}}\right) - \textrm{arcsinh}(u) = \pm\sqrt{2\vert\sigma\vert} (\xi - \xi_{0})\, .
 \label{eq:darkcssolution}
\end{equation}
The obtained solution corresponds to the heteroclinic orbit of Fig.~\ref{fig:bdg}(b). Black solitary profiles given by Eq.~(\ref{eq:darkcssolution}) are shown in Fig.~\ref{fig:DarkCentrosymm}(a). In the limit $\vert u\vert\leq u_{\infty}\ll 1$ we recover from Eq.~(\ref{eq:darkcssolution}) the expression $u(\xi)= \pm u_{\infty} \textrm{tanh}[u_{\infty}\sqrt{ 2\sigma}\,(\xi - \xi_{0})]$ corresponding to the one-dimensional black soliton of the cubic NSE. Moreover, the dependence of the FWHM, $\Delta\xi$, as a function of the amplitude at infinity $u_{\infty}$ can be easily obtained from Eq.~(\ref{eq:darkcssolution}) by setting $u^{2}(\Delta\xi/2)=u_{\infty}^{2}/2$, and reads as
\begin{equation}
\Delta\xi (u_{\infty}) = \sqrt{\frac{2}{\vert\sigma\vert}}\left[\frac{\sqrt{1+u_{\infty}^{2}}}{u_{\infty}}\,\textrm{arctanh}\left( \sqrt{\frac{1+u_{\infty}^{2}}{2+u_{\infty}^{2}}}\,\right)-\textrm{arcsinh} \left( \frac{u_{\infty}}{\sqrt{2}}\right)\!\right] \! . 
\label{eq:darkcsexistcurv}
\end{equation}
As depicted in Fig.~\ref{fig:DarkCentrosymm}(b), the amplitude bistability property of bright solitary waves is absent here. For increasing values of $u_{\infty}$, the FWHM tends to the nonzero constant $\ln (2)/\sqrt{2\vert\sigma\vert}$. 
\par
Regarding the stability of the found black solitary waves, they are all stable since the saturable nonlinearity~(\ref{eq:quadraticsaturable}) belongs to the class of local nonlinearities for which the resulting solutions are stable for any velocity $v$~\cite{Chen96}.

\subsection{Grey Solitary Waves}
\begin{figure}
\begin{center}
\hspace*{-0.3cm}
\hbox{\vbox{\vskip -0.0cm \epsfig{figure=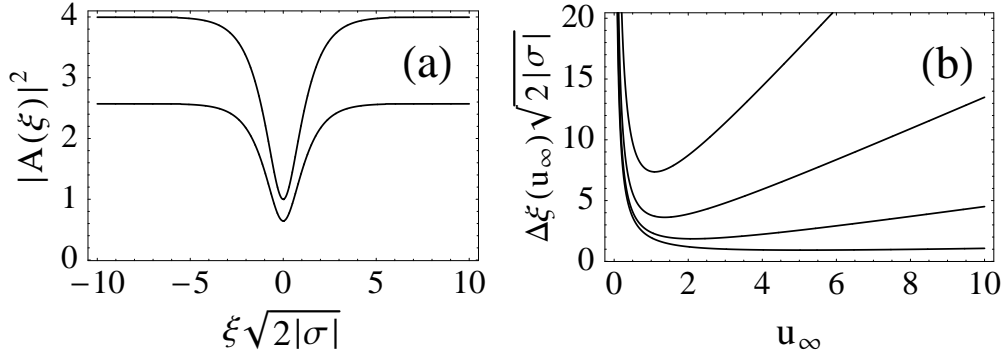, scale=1.04}}}
\end{center}
\vspace*{-0.2cm}
\caption[\bf]{\small (a) Transverse profiles of grey solitary waves for $u_{\infty}^{2}=4$ and $u_{0}^{2}=1$ (upper curve), and $u_{\infty}^{2}=2.6$ and $u_{0}^{2}=0.64$ (lower curve), as obtained from Eq.~(\ref{eq:greycssolution}). (b) Existence curves for grey solitary waves, where the FWHM $\Delta\xi$ is plotted versus the amplitude at infinity $u_{\infty}$ for $u_{0}/u_{\infty}=0.1,0.4,0.7$, and $0.9$ (lower to upper curves).}
\label{fig:GreyCentrosymm}
\end{figure}

Fundamental grey solitary waves fulfill $u_{0}\neq0$, $u_{\infty}\neq0$, and $u_{0}'=0$. From Eqs.~(\ref{eq:greypropgconst}),~(\ref{eq:greylo}), ~(\ref{eq:greysolitarywaves}), and~(\ref{eq:quadraticsaturable}) we find that $\Gamma=\sigma/(1+u_{0}^{2})$ and $l_{0}^{2}=(2\sigma u_{0}^{2}\,u_{\infty}^{4})/(1+u_{0}^{2})$. This last expression imposes that $\sigma>0$. The energy equation is given now by
\begin{equation}
\left( \frac{du}{d\xi}\right)^{2} = \frac{2\sigma\,(u^{2}-u_{0}^{2})(u_{\infty}^{2}-u^{2})^{2}}{(1+u^{2})(1+u_{0}^{2})\,u^{2}}\, . \label{eq:greycsweqint2}
\end{equation}
Notice that the structure of Eq.~(\ref{eq:greycsweqint2}) precludes the existence of solutions satisfying $u_{\infty}^{2}\leq u^{2}\leq u_{0}^{2}$. This means that there cannot be {\em hump-like} waves on a constant nonzero background; and so we look for {\em dip-like} waves (i.e. $u_{0}^{2}\leq u^{2}\leq u_{\infty}^{2}$) in a constant nonzero background. Upon integration of Eq.~(\ref{eq:greycsweqint2}) we derive the following implicit relation for the envelope distribution $u(\xi)$
\begin{eqnarray}
\sqrt{\frac{1+u_{\infty}^{2}}{u_{\infty}^{2}-u_{0}^{2}}} \,\textrm{arctanh}\left(\sqrt{\frac{(1+u_{\infty}^{2})(u^{2}-u_{0}^{2})}{(1+u^{2})(u_{\infty}^{2}-u_{0}^{2})}}\,\right) &-& \textrm{arcsinh}\left(\sqrt{\frac{u^{2}-u_{0}^{2}}{1+u_{0}^{2}}}\,\right)\nonumber\\
&=& \pm\sqrt{\frac{2\vert\sigma\vert}{1+u_{0}^{2}}}\, (\xi - \xi_{0})\, .
\label{eq:greycssolution}
\end{eqnarray}
The obtained solution corresponds to the homoclinic orbits of Fig.~\ref{fig:bdg}(c). Figure~\ref{fig:GreyCentrosymm}(a) illustrates the profiles of two grey solitary waves having the same FWHM. The existence properties of grey-type solutions are described by an {\em existence surface}, which can be easily found from Eq.~(\ref{eq:greycssolution}) by setting $u^{2}(\Delta\xi)=(u_{0}^{2}+u_{\infty}^{2})/2$. Figure~\ref{fig:GreyCentrosymm}(b) represents four slices of this surface for various ratios of $u_{0}/u_{\infty}$ when varying $u_{\infty}$. The amplitude bistability feature depends now on this ratio and disappears as $u_{0}/u_{\infty}\to 0$. Moreover, in contrast with bright and black solutions, the phase structure of grey solitary waves is nontrivial. The term $\int^{\xi}d\tau/u^{2}(\tau)$ in Eq.~(\ref{eq:phi2}) can be calculated exactly by resorting to Eq.~(\ref{eq:greycsweqint2}) together with a method similar to the one employed to evaluate the power $P_{s}$ of bright solitary waves [see Eq.~(\ref{eq:calculationPs})]. The resulting analytical form of the phase is
\begin{multline}
\phi(\xi,\zeta) = \phi_{0}+\left( \frac{\sigma}{1+u_{0}^2} + \frac{v^{2}}{2}\right)\!\zeta + v\xi - \textrm{arcsin}\left(\frac{u_{0}}{u(\xi)}\sqrt{\frac{1+u^{2}(\xi)}{1+u_{0}^2}}\,\right)\\
+ u_{0}\sqrt{\frac{1+u_{\infty}^{2}}{u_{\infty}^{2}-u_{0}^{2}}}\,\textrm{arctanh}\left(\sqrt{\frac{(1+u_{\infty}^{2})[u^{2}(\xi)-u_{0}^{2}]}{(u_{\infty}^{2}-u_{0}^{2})[1+u^{2}(\xi)]}}\,\right), 
\label{eq:greyphi}
\end{multline}
where $u=u(\xi)$ is given implicitly by Eq.~(\ref{eq:greycssolution}). In the limit $u_{0}^{2}\leq u^{2} \leq u_{\infty}^{2}\ll1$, one retrieves from Eq.~(\ref{eq:greycssolution}) the profile for the one-dimensional grey soliton in a nonlinear cubic medium, which reads as
\begin{eqnarray}
u(\xi)=\pm\sqrt{u_{\infty}^{2}-(u_{\infty}^{2}-u_{0}^{2})\,\textrm{sech}^{2}\left[\sqrt{2\vert\sigma\vert(u_{\infty}^{2}-u_{0}^{2})}\,(\xi - \xi_{0}) \right]}\, ,
 \label{eq:greycubicsolution}
\end{eqnarray}
and the corresponding phase dependence follows straight away from Eq.~(\ref{eq:greyphi}).
\par
With respect to the the stability of the found grey solitary waves, it is an open problem to determine the regions defined by $u_{0}$ and $u_{\infty}$ for which these solutions are stable. This issue will be addressed in a subsequent work.
\par
\section{Conclusions}
In this paper we have constructed explicit analytical expressions for solitary wave solutions of the nonlinear Schr\"odinger equation with a saturable nonlinearity 
 of the form $\propto (1+\vert A\vert^{2})^{-2}$ arising, for instance, in the propagation of nonlinear optical beams in centrosymmetric photorefractive materials. The bright and black soliton solutions had only been found numerically previously in the context of optical applications of the model equation.  In the case of bright solitons solutions we have also performed a study of their stability. The analysis of the stability of grey solitons is not so simple and could be the goal of future research in this area.
\par
Our work can be extended in several directions. The first one, by considering more complicated saturable nonlinearities such as those arising in media with electromagnetically induced transparency \cite{Light}. Secondly, by studying saturable vector media with several fields involved. Finally, another interesting extension could be the analysis of the model with space-dependent parameters, resorting to the technique of Lie symmetries~\cite{Nuestro,Nuestro2}. 

\par 

\section*{Acknowledgements}
We would like to thank I. A. Molotkov for discussions.
This work has been partially supported by grants  FIS2006-04190 (MEC), and  PCI-08-0093 (Junta de Comunidades de Castilla-La Mancha, Spain).
\par


\begin{thebibliography}{99}

\bibitem{Vazquez} V{\'a}zquez L, Streit L, P{\'e}rez-Garc{\'\i}a V M, Eds. Nonlinear Klein-Gordon and Schr{\"o}dinger systems: Theory and Applications. Singapur: World Scientific;  1997.

\bibitem{Peyrard} Dauxois T, Peyrard M. Physics of Solitons. Cambridge: Cambridge University Press; 2006.

\bibitem{Soler} Brezzi F, Markowich P A. The three-dimensional Wigner-Poisson problem: existence, uniqueness and approximation. Math Mod Meth Appl Sci 1991;14:35-61.

\bibitem{Soler2} L{\'o}pez J L, Soler J. Asymptotic behaviour to the 3D Schr{\"o}dinger/Hartree-Poisson and Wigner-Poisson systems. Math Mod Meth Appl Sci  2000;10:923-943.

\bibitem{Kivshar} Kivshar Y, Agrawal G P. Optical Solitons: From fibers to Photonic crystals. New York: Academic Press; 2003.

\bibitem{Hasegawa} Hasegawa A.  Optical Solitons in Fibers. Berlin: Springer-Verlag; 1989.

\bibitem{Dodd} Dodd R K, Eilbeck J C, Gibbon J D, Morris H C. Solitons and nonlinear wave equations. New York:  Academic Press; 1982.

\bibitem{fundamentals} Rosales J L,  S{\'a}nchez-G{\'o}mez J L. Nonlinear Sch{\"o}dinger equation
coming from the action of the particles gravitational field on the quantum potential. Phys Lett A  1992; 66:111-115.

\bibitem{Fedele} Fedele R, Miele G, Palumbo L, Vaccaro V G. Thermal wave model for nonlinear longitudinal dynamics in particle accelerators. Phys Lett A 1993;173:407-413.

\bibitem{Dalfovo} Dalfovo F, Giorgini S, Pitaevskii L P, Stringari S. Theory of Bose-Einstein condensation in trapped gases. Rev Mod Phys 1999;71:463-512.

\bibitem{Davidov} Davydov A S. Solitons in Molecular Systems. Dordrecht: Reidel; 1985.

\bibitem{Scott} Scott A. Nonlinear Science: Emergence and dynamics of coherent structures. Oxford: Oxford Appl and Eng Mathematics Vol. 1; 1999.

\bibitem{Zakharov} Zaharov V E, L'vov V S, Starobinets S S. Spin-wave turbulence beyond the parametric excitation threshold. Sov Phys Usp  1975;17:896-919.

\bibitem{Sulembook} Sulem C, Sulem P. The nonlinear Schr{\"o}dinger equation: Self-focusing and wave collapse. Berlin: Springer; 2000.

\bibitem{SIAMFibich} Fibich G, Papanicolau G. Self-focusing in the perturbed and unperturbed
 nonlinear Schr{\"o}dinger equation in critical dimension. SIAM J Appl Math 1999;60:183-240.

\bibitem{Yu} Yu M Y, Shukla P K, Spatschek K H. Localization of high-power laser pulses in plasmas. Phys. Rev. A 1978;18:1591-1596.

\bibitem{Marburger} Marburger J H, Dawes E. Dynamical formation of a small-scale filament. Phys Rev Lett 1968;21:556-558.

\bibitem{Herrmann} Gatz S, Herrmann J. Soliton propagation in materials with saturable nonlinearity. J Opt Soc Am B 1991;8:2296-2302.

\bibitem{Krolikowski93} Krolikowski W, Luther-Davis B. Dark optical solitons in saturable nonlinear media. Opt Lett 1993;18:188-90.

\bibitem{Segev94} Segev M, Valley G C, Crosignani B, Porto P D, Yariv A. Steady-state spatial screening solitons in photorefractive materials with external applied field. Phys Rev Lett 1994;73: 3211.

\bibitem{Christodoulides95}  Christodoulides D N, Carvalho M I. Bright, dark, and gray spatial soliton states in photorefractive media. J Opt Soc Am B 1995;12:1628-1633. 
 
\bibitem{Segev96} Segev M, Shih MF, Valley GC. Photorefractive screening solitons of high and low intensity.  J Opt Soc Am B 1996;13:706-18.

\bibitem{Krolikowski00} Krolikowski W, Ostrovskaya E A, Weilnau C, Geisser M, McCarthy G, Kivshar Y S, Denz C, Luther-Davis B. Observation of dipole-mode vector solitons. Phys Rev Lett 2000;85:1424-1427.

\bibitem{Calvo02} Calvo G F, Sturman B, Agull\'{o}-L\'{o}pez F, Carrascosa M. Solitonlike beam propagation along light-induced singularity of space charge in fast photorefractive media. Phys Rev Lett 2002;89:033902.

\bibitem{Shumelyuk} Shumelyuk A, Shcherbin K, Odoulov S, Sturman B, Podivilov E, Buse K. Slowing down of light in photorefractive crystals with beam intensity coupling reduced to zero. Phys Rev Lett 2004;93:243604.

\bibitem{Quirino} Garcia-Quirino G S, Sanchez-Mondragon J J, Stepanov S. Nonlinear surface optical waves in photorefractive crystals with a diffusion mechanism of nonlinearity. Phys Rev A 1995;51:1571-1577.

\bibitem{Mamaev} Mamaev A V, Saffman M. Selection of unstable patterns and control of optical turbulence by fourier plane filtering. Phys Rev Lett 1998;80:3499-3502.

\bibitem{Calvo00} Calvo G F, Sturman B, Agull\'{o}-L\'{o}pez F, Carrascosa M. Singular behavior of light-induced space charge in photorefractive media under an ac field. Phys Rev Lett 2000;84:3839-3842.

\bibitem{Podivilov} Podivilov E V, Sturman B I, Pedersen H C, Johansen P M. Critical enhancement of photorefractive beam coupling. Phys Rev Lett 2001;85:1867-1870.

\bibitem{Carvalho07}  Carvalho M I, Fac\~{a}o M, Christodoulides D N. Self-bending of dark and gray photorefractive solitons. Phys Rev E 2007;76:016602. 

\bibitem{SegevAgranat} Segev M and Agranat A J. Spatial solitons in centrosymmetric photorefractive media. Opt Lett 1997;22:1299-1301.

\bibitem{KLTN1} DelRe E, Crosignani B, Tamburrini M, Segev M, Mitchell M, Refaeli E, and Agranat A J. One-dimensional steady-state photorefractive spatial solitons in  centrosymmetric paraelectric potassium lithium tantalate niobate. Opt Lett  1998;23: 421-423.

\bibitem{KLTN2} DelRe E, Tamburrini M, Segev M, Refaeli E, and Agranat A J. Two-dimensional photorefractive spatial solitons in centrosymmetric paraelectric potassium-lithium-tantalate-niobate. Appl Phys Lett 1998;73:16-18.

\bibitem{CSF1} Li B, Chen Y. On exact solutions of the nonlinear Schr\"odinger equations in optical fiber. Chaos, Solitons and Fractals 2004;21:241-247.

\bibitem{CSF2} Zayed E M E, Zedan H E. On the solution of the nonlinear Schr\"odinger equation. Chaos, Solitons and Fractals 2003;16:133-145.

  \bibitem{Nuestro} Belmonte-Beitia J, P\'erez-Garc\'ia V M, Vekslerchik V, Torres P J. Lie symmetries and solitons in nonlinear systems with spatially inhomogeneous nonlinearities. Phys Rev Lett 2007;98:064102.

 \bibitem{Nuestro2} Belmonte-Beitia J, P\'erez-Garc\'ia V M, Vekslerchik V, Torres P J. Lie symmetries, qualitative analysis and exact solutions of nonlinear Schr\"odinger equations with inhomogeneous nonlinearities. Discrete and Continuous Dynamical Systems-B  2008;9:221-233.

\bibitem{CSF3} Zhu J M, Ma Z Y. Exact solutions for the cubic-quintic nonlinear Schr\"odinger equation. Chaos, Solitons and Fractals  2007;33:958-964.

\bibitem{CSF4} Zhu S D. Exact solutions for the high-order dispersive cubic-quintic nonlinear Schr\"odinger equation by the extended hyperbolic auxiliary equation method. Chaos, Solitons and Fractals  2007;34:1608-1612.

\bibitem{JVV} Belmonte-Beitia J, P\'erez-Garc\'ia V M, Vekslerchik V. Modulational instability, solitons and periodic waves in models of quantum degenerate Boson-Fermion mixtures. Chaos, Solitons and Fractals  2007;32:1268-1277. 

\bibitem{VK} Vakhitov M G, Kolokolov A A. Stationary solutions of the wave equation in a medium with nonlinearity saturation. Radiophys. and Quantum Electronics 1973;16:783-789.

\bibitem{Laedke} Laedke E W, Spatschek K H, Stenflo L. Evolution theorem for a class of perturbed envelope soliton solutions. J Math Phys 1983;24:2764-2769.

\bibitem{Chen96} Chen Y. Stability of black solitons in media with arbitrary nonlinearity. Opt Lett 1996;21:462-64.

\bibitem{Light} Michinel H, Paz-Alonso M, P\'erez-Garc\'{\i}a V M, Turning Light into a Liquid via Atomic Coherence. Phys. Rev. Lett. 2006;96:023903.

\end{thebibliography}
\end{document}